\documentclass[prd,aps,preprint,tightenlines]{revtex4}
\usepackage{epsfig}
\begin{document}
\voffset=2.5cm
\preprint{UMD PP\#01-055}
\preprint{DOE/ER/40762-230}
\title{Large-$N_c$ Quark Distributions in the Delta \\ and Chiral
Logarithms in Quark Distributions of the Nucleon} 

\author{Jiunn-Wei Chen}
\email{jwchen@physics.umd.edu}
\affiliation{Department of Physics, 
University of Maryland, 
College Park, Maryland 20742 }
\author{Xiangdong Ji}
\email{xji@physics.umd.edu}
\affiliation{Department of Physics, 
University of Maryland, 
College Park, Maryland 20742 }

\vspace{0.2in}
\date{\today}

\vspace{0.2in}
\begin{abstract}

In a world with two quark flavors and a large number of 
colors ($N_c$), the 
polarized and unpolarized quark distributions in the delta 
are completely determined by those in the nucleon 
up to relative ${\cal O}(1/N_c^2)$. In particular, we find 
$q_{\Delta}(x)
=\left[(1\pm 2T_z)u_N(x)+ (1\mp 2T_z)d_N(x)\right]/2\left(1+ 
{\cal O}(1/N_c^2)\right)$ 
and $\Delta q_\Delta(x) =\left[(5\pm 2T_z)\Delta u_N(x) 
+ (5\mp 2T_z)\Delta d_N(x)\right]/10\left(1
+ {\cal O}(1/N_c^2)\right)$, where $q=u, d$ and 
$T_z$ the charge state of a delta. The result can be used to 
estimate the leading chiral-logarithmic corrections to the 
quark distributions in the nucleon.

\end{abstract}
\maketitle

Among baryon resonances, the delta  
is very special: In a world with 
two quark flavors and a large number of colors $(N_c)$, 
consistency conditions for meson-baryon scattering imply
that there is a (nearly) degenerate hadron multiplet 
with quantum numbers $J = T = 1/2,~ 3/2, ~..., ~N_c/2$ \cite{DM1,GS}. 
In the real world of $N_c=3$, the delta resonance 
and the nucleon can be identified with this multiplet. 
Hence many properties of the delta 
in the limit of $N_c\rightarrow \infty$ can be 
related to those of the nucleon by simple Clebsch-Gordon
coefficients. Many of these relations, particularly those
accurate to order $1/N_c^2$, are shown to be valid in data 
at 10\% level \cite{DM1,Jenkins,Jenkins1}.

In this paper, we explore the large-$N_c$ constraint
on the quark distributions in the delta. We show that 
the unpolarized up and down quark distributions in 
the delta ($\Delta$) are related to those in the nucleon $(N)$ by 
\begin{equation}
 q_{\Delta}(x)
={1\over 2}\left[(1\pm 2T_z)u_N(x)
+ (1\mp 2T_z)d_N(x)\right]\left(1+ {\cal O}(1/N_c^2)\right) \ ,
\end{equation}
where $q=u, d$, and $T_z$ specifies the charge state 
of a delta. In particular, the distributions in the 
$\Delta^+$ and the nucleon ($N$) are the same. Since 
these distributions are of order $N_c$,  
their ratios in $\Delta^+$ and $N$ are equal to 1 
up to corrections of order $1/N_c^2$. 
Similar relations are found for the 
quark helicity and transversity distributions \cite{filippone}:
\begin{eqnarray}
 \Delta q_\Delta(x)
&=& {1\over 10}\left[(5\pm 2T_z)\Delta u_N(x)+ (5\mp 2T_z)\Delta
d_N(x)\right]\left(1 
+ {\cal O}(1/N_c^2)\right) \ , \nonumber \\ 
 \delta q_\Delta(x)
&=& {1\over 10}\left[(5\pm 2T_z)\delta u_N(x) + (5\mp 2T_z)\delta
d_N(x)\right] \left(1
+ {\cal O}(1/N_c^2)\right) \ . 
\end{eqnarray}
Using the above relations, we can make estimates
of the delta contributions to the leading chiral-logarithms 
in the quark distributions of the nucleon \cite{others,CJ,AS}. 
The interplay between chiral behavior and large $N_c$ quantities 
is reminiscent of what is seen for static baryon properties \cite{tom}.
The quark distributions in the delta, particularly the nucleon-delta
transition distribution:  
\begin{equation}
 \Delta u_{N\Delta}(x)-\Delta d_{N\Delta}(x) = \sqrt{2} (\Delta u_N(x)-\Delta d_N(x))\left(1 
   + {\cal O}(1/N_c^2)\right)
\end{equation}
can be used to understand the 
parton distributions in nuclei \cite{thomas}. 
Some large $N_c$ results about the  
$N-\Delta$ transition off-forward distributions 
were discussed in Ref. \cite{frankfurt}.
Finally, 
the $\Delta$ distributions can be calculated on a 
lattice when the quark masses are sufficiently large 
and a stable delta exists. 

In the large-$N_c$ limit, all baryons including the nucleon
and the delta are infinitely heavy. For simplicity we assume 
they have a four-velocity $v^\mu$ and, without loss of generality,
we take $v^\mu=(1,0,0,0)$. The spin-flavor wave function
of a baryon state $|J=T, J_z, T_z\rangle$ can be taken as a 
product of the complete symmetric tensors 
$\psi^{(\alpha_1\alpha_2...\alpha_{2J})}\sim \psi_{JJ_z}$ 
and $\chi^{(a_1a_2...a_{2J})}\sim \chi_{JT_z}$ in the spin 
and flavor spaces, respectively. 
Introduce the following twist-two quark operators,
\begin{eqnarray}
O_{q}^{\mu _{1}\cdots \mu _{n}} &=&\overline{q}\gamma ^{(\mu
_{1}}iD^{\mu _{2}}\cdots iD^{\mu _{n})}q \  , \nonumber \\
O_{\Delta q}^{\mu _{1}\cdots \mu _{n}} &=&\overline{q}\gamma
^{(\mu _{1}}\gamma _{5}iD^{\mu _{2}}\cdots iD^{\mu _{n})}q  \ , 
\nonumber \\
O_{\delta q}^{\alpha\mu _{1}\cdots \mu _{n}} 
&=&\overline{q}\sigma
^{[\alpha(\mu _{1}]}\gamma_5iD^{\mu _{2}}\cdots iD^{\mu _{n})}q   \ , 
\label{op}
\end{eqnarray}
where $(\cdots)$ and $[\cdots]$ denote, respectively, 
the symmetrization and antisymmetrization of the indices in between,
{\it and} trace subtractions. The above operators can be
made dimensionless by dividing $n+2$ powers of a baryon mass.
[Since the nucleon and delta masses differ at order $1/N_c$,
one can take either mass without affecting the result.]
Considering those components which survive the non-relativistic
limit, we define
\begin{eqnarray}
    {\cal O}_q^{(n)} &&= O^{(00\cdots 0)}_qN_c^3/M^{n+2} \ , \nonumber \\
    {\cal O}_{\Delta q}^{(n)i} &&= O^{(i0\cdots 0)
   }_{\Delta q}N_c^3/M^{n+2} \ , \nonumber \\
 {\cal O}_{\delta q}^{(n)i} &&= O^{(i0\cdots 0)
   }_{\delta q}N_c^3/M^{n+2} \ . 
\end{eqnarray}
The matrix elements of the above operators in the baryon 
multiplet are 
\begin{eqnarray}
\left\langle J,J_z,T_z\left| {\cal O}_{q}^{(n)}\right| J,J_z,T_z\right\rangle &=&
 \left\langle x^{n-1}\right\rangle_{q}
   v^{(0}\cdots v^{0)} \ ,  \nonumber \\
\left\langle J, J_z, T_z\left| {\cal O}
_{\Delta q}^{(n)i}\right|J,J_z,T_z\right\rangle
  &= & \left\langle x^{n-1}\right\rangle _{\Delta q}
  \psi_{JJ_z}^\dagger J^{(i}v^0\cdots v^{0)}\psi_{JJ_z}   \ , 
 \nonumber \\
\left\langle J, J_z,T_z\left| {\cal O}
_{\delta q}^{(n)i}\right| J, J_z, T_z\right\rangle
  &=& \left\langle x^{n-1}\right\rangle_{\delta q}
\psi_{JJ_z}^\dagger J^{[i}v^{(0]}v^0\cdots v^{0)} \psi_{JJ_z}  
  \ . 
\end{eqnarray}
The coefficients in front of the various structures 
are the moments of the unpolarized and polarized 
quark distributions: 
\begin{eqnarray}
\left\langle x^{n-1}\right\rangle _{q} &=&\int_{0}^{1}dxx^{n-1}\left(
q\left( x\right) +\left( -1\right) ^{n}\overline{q}\left(
x \right) \right)  \ , 
     \nonumber \\
\left\langle x^{n-1}\right\rangle _{\Delta q} &=&\int_{0}^{1}dxx^{n-1}\left(
\Delta q\left( x\right) +\left( -1\right) ^{n-1}\Delta \overline{q%
}\left( x\right) \right) \ , 
    \nonumber \\
\left\langle x^{n-1}\right\rangle _{\delta q} &=&\int_{0}^{1}dxx^{n-1}\left(
\delta q\left( x\right) +\left( -1\right) ^{n}\delta \overline{q%
}\left( x\right) \right) \ , 
    \nonumber \\
 &&~~~~~~~~~(n=1,2,3,...) 
\end{eqnarray}
The isospin dependence of the above distributions is implicit. 
When we refer to the quark distributions in the nucleon,
we mean those in the proton $(T_z=1/2)$. For the delta, unless 
stated otherwise, we mean the +1 charge state, i.e., 
$\Delta^+$ with $T_z=1/2$. 

We now consider quark distributions in the delta separately for
different spin-isospin channels:

\vspace{0.2in}
\noindent
1. {\bf Scalar-isoscalar channel}:

Consider the unpolarized isoscalar quark distribution $u(x)+d(x)$. Its
moments are defined as the matrix elements of the 
twist-two operators ${\cal O}^{(n)}_{u+d}$. 
Expressed in terms of quark and gluon fields, the operators
are sums of one- and many-body quark-gluon operators. 
When inserted in between baryon states, 
all contributions are of order $N_c\cdot N_c^{1-n}$ (
where the second factor comes from the trivial normalization
factor in Eq. (5)). Therefore, 
in the limit of $N_c\rightarrow \infty$, $u(x)+d(x)$ scales as 
$N_c^2\phi(xN_c)$, where $\phi(x)$ is an $N_c$ independent 
function \cite{polyakov}. 

All operators ${\cal O}^{(n)}_{u+d}$ are spin and 
isospin-independent. Therefore, like the baryon masses, 
they can be expanded in terms of the angular momentum 
operator $J$ \cite{Jenkins}
\begin{equation}
      {\cal O}_{u+d}^{(n)}/N_c^{2-n}
    = a_0^{(n)} + a_1^{(n)} \left({J\over N_c}\right)^2 + ... \ , 
\end{equation}
where $a_0^{(n)}$ and $a_1^{(n)}$ are constants of order 1
in $N_c$ counting. Here we have neglected possible 
contributions from 
isospin violation. The matrix elements of the above 
operators in the nucleon ($J=1/2$) and the delta ($J=3/2$) are
the same up to corrections of relative order $(1/N_c)^2$. 
We therefore conclude that
the $u(x)+d(x)$ distributions in the nucleon and delta 
are related through
\begin{equation}
     {u_N(x)+ d_N(x)\over u_\Delta(x)+ d_\Delta(x)}
  = 1 + {\cal O}(1/N_c^2) \ . 
\label{ss}
\end{equation}
The above relations are true for any charged state of the
nucleon and delta.

The result has an immediate application. 
The leading chiral logarithms 
of the quark distributions in the nucleon 
can be calculated in heavy-baryon chiral perturbation theory 
\cite{CJ,AS}. The generalization to large-$N_c$ chiral
perturbation theory is straightforward. In this expansion, 
$m_\pi\rightarrow 0$, $N_c\rightarrow \infty$, and $m_\pi N_c$
fixed. The delta contribution to the leading chiral behavior of the moments of the 
$u(x)+d(x)$ distribution in the nucleon
was calculated in 
Ref. \cite{AS}. The result is 
\begin{equation}
      \delta\langle x^{n-1}_{(u+d)N} \rangle_{{\rm from} ~\Delta} = 
     -4{g_{\pi N\Delta}^2\over (4\pi f_\pi)^2} J_1(\Delta, m_\pi)
        \left(\langle x_{(u+d)N}^{n-1}\rangle^0 -\langle x_{(u+d)
        \Delta}^{n-1}\rangle^0\right) \ , 
\end{equation}
where 
\begin{equation}
  J_1(\Delta, m_\pi)
  = (m_\pi^2-2\Delta^2)\log\left({m_\pi^2\over
    \mu^2}\right)
  + 2\Delta\sqrt{\Delta^2-m_\pi^2}
\log\left({\Delta-\sqrt{\Delta^2-m_\pi^2+i\epsilon}
       \over \Delta + \sqrt{\Delta^2- m_\pi^2+i\epsilon}}
\right) \ , 
\end{equation}
and $g_{\pi N\Delta}$ is the $\pi$-$N$-$\Delta$ coupling,  
and $\Delta$ is the delta-nucleon mass difference. The superscript $0$
labels the moments in the chiral limit.
Since $g_{\pi N\Delta}\sim N_c$, $f_\pi\sim \sqrt{N_c}$,
$\langle x_{(u+d)N}^{n-1}\rangle^0 -\langle x_{(u+d)
        \Delta}^{n-1}\rangle^0 \sim 1/N_c^{-n}$, the
above correction is of order $N_c^{1-n}$ in $N_c$ counting. 
Hence the chiral corrections are subleading both 
in $m_\pi/4\pi f_\pi$ and in $1/N_c$. 

\vspace{0.2in}
\noindent
2. {\bf Scalar-isovector channel}:

The unpolarized isovector distribution $u(x)-d(x)$ is defined 
in terms of the twist-two operators ${\cal O}_{u-d}^{(n)}$. 
Because of the cancellation from the up and down quark contributions,
the matrix elements in the nucleon and delta states
are of order $N_c^{1-n}$ in $N_c$ counting. These operators 
can be expanded as \cite{Jenkins},
\begin{equation}
     {\cal O}_{u-d}^{(n)}/N_c^{1-n}  = a_0^{(n)}T^3 + a_1^{(n)} 
         {J^i G^{i3}\over N_c} + a_2^{(n)}{J^2T^3\over N_c^2}
      + ... \ , 
\end{equation}
where $G^{ia}$ are generators of spin-flavor
SU(4). For two flavors, $4J_iG^{ia}=(N_c+2)T^a$. \cite{Jenkins1}
From the above expansion, we conclude that the
$u-d$ distributions in the proton and the $\Delta^+$ are
the same up to order $1/N_c^2$, 
\begin{equation}
      {u_N(x) - d_N(x)\over u_{\Delta^+}(x)-d_{\Delta^+}(x)}
   =1 + {\cal O}(1/N_c^2) \ . 
\label{sv}
\end{equation}

Combining the relations in Eqs. (\ref{ss}) and (\ref{sv}), we have
\begin{eqnarray}
          u_\Delta(x) &=& \left\{\left({1\over 2}+T_z\right)u_N(x)
                      + \left({1\over 2}-T_z\right)d_N(x)\right\}\left(1 + 
{\cal O}(1/N_c^2)\right)  \ , \nonumber \\
  d_\Delta(x) &=& \left\{\left({1\over 2}-T_z\right)u_N(x)
                      + \left({1\over 2}+T_z\right)d_N(x)\right\}\left(1 
+ {\cal O}(1/N_c^2)\right) \ . 
\end{eqnarray}
Here we have allowed the delta in a general
charge state, $T_z$. 

Consider now the chiral logarithms in the $u(x)-d(x)$
distribution in the nucleon. The delta contribution
is $(n>1)$, \cite{AS}
\begin{equation}
\delta\langle x^{n-1}_{(u-d)N} \rangle_{{\rm from} ~\Delta}
= 
     -{4g_{\pi N\Delta}^2\over (4\pi f_\pi)^2} J_1(\Delta, m_\pi)
        \left(\langle x_{(u-d)N}^{n-1}\rangle^0 -{5\over 3}
          \langle x_{(u-d)
        \Delta^+}^{n-1}\rangle^0\right) \ . 
\label{delta}
\end{equation}
To make an estimate, we assume the large-$N_c$ relations,
\begin{eqnarray}
         g_{\pi N \Delta} &=& {3\over 2\sqrt{2}}g_A \ , 
          \nonumber \\
 \langle x_{(u-d)N}^{n-1}\rangle^0 &=& 
          \langle x_{(u-d)
        \Delta^+}^{n-1}\rangle^0 \ . 
\end{eqnarray}
Then the contribution from Eq. (\ref{delta}) plus that from the 
$\pi$-$N$ loop \cite{CJ, AS} yields,
\begin{equation}
  \delta\langle x^{n-1}_{(u-d)N} \rangle
= C_n {1\over (4\pi f_\pi)^2}
  \left[3g_A^2 J_1(\Delta,m_\pi) - (3g_A^2+1)m_\pi^2\log\left({m_\pi^2\over \mu^2}
   \right)\right] \ , 
\end{equation}
where $C_n = \langle x^{n-1}_{(u-d)N} \rangle^0$. 
Each term in the square bracket is of order $N_c^2$; however,
the nucleon and delta contributions cancel up to and including 
order $N_c$, where the large $N_c$ limit is taken with fixed $m_\pi$. 
The remaining contribution is of order $N_c^0$; and hence
the chiral correction is again subleading in $1/N_c$. 
In the real world, the delta contribution has the same sign 
as the nucleon contribution and is about twice in magnitude.
Of course, the analytical contributions there might not be 
so small compared with the non-analytical ones considered here. 

\vspace{0.2in}
\noindent
3. {\bf Vector-isovector channel}:

Now we turn to the quark helicity distributions in the delta.
First, let us consider the isovector distribution $\Delta u(x)-\Delta d(x)$. 
The corresponding twist-two operators
${\cal O}^{(n)i}_{\Delta u-\Delta d}$ is proportional 
to the vector-isovector operator $J^iT^a$. Thus, 
in the large-$N_c$ limit, their matrix elements are 
proportional to $N_c\cdot N_c^{1-n}$. Using the same arguments  
for the pion-baryon coupling or isovector
magnetic moment \cite{DM1,Jenkins1}, we find that the ratio
of the distribution in the delta and nucleon
is that of the spin-isospin
Clebsch-Gordon coefficients, 
\begin{equation}
        {\Delta u_\Delta(x)-\Delta d_\Delta(x)\over  
       \Delta u_N(x)-\Delta d_N(x)}
   = {2T_z\over 5} + {\cal O}(1/N_c^2) \ .
\label{vv}
\end{equation}
where again $T_z$ is the isospin projection of the delta. 
In particular, for the $\Delta^+$ state, the ratio is 1/5. 
Similarly, we have the relation for the transversity
distribution
\begin{equation}
        {\delta u_\Delta(x)-\delta d_\Delta(x)\over  
       \delta u_N(x)-\delta d_N(x)}
   = {2T_z\over 5} + {\cal O}(1/N_c^2) \ .
\end{equation}

To find the delta contribution to the leading
chiral logarithms in $\Delta u_N(x)-\Delta d_N(x)$, 
we need to define the spin-dependent
delta-nucleon transition distribution, 
$\Delta q_{N\Delta}(x)$ \cite{thomas}.
Its moments are defined by the matrix elements,
\begin{equation}
  \left\langle {3\over 2},J_z',{1\over 2}\left|
      {\cal O}^{(n)i}_{\Delta u-\Delta d}\right|
  {1\over 2},J_z, {1\over 2} \right\rangle
  = \langle x^{n-1}\rangle_{(\Delta u-\Delta d)N\Delta} 
    \psi^\dagger_{{3\over 2}J_z'} J^{(i}v^0\cdots v^{0)}\psi_{{1\over 2}J_z}  \ . 
\end{equation} 
The large-$N_c$ constraint yields,
\begin{equation}
       \langle x^{n-1} \rangle_{(\Delta u-\Delta d)N\Delta} 
        = \sqrt{2} \langle x^{n-1} \rangle_{(\Delta u-\Delta d)N}
\left(1  + {\cal O}(1/N_c^2)\right) \ ,   
\end{equation}
where $\sqrt{2}$ is a ratio of Clebsch-Gordon coefficients.
This result can be compared with the quark model one
in Ref. \cite{thomas}.

\begin{figure}[t]
\begin{center}
\epsfig{figure=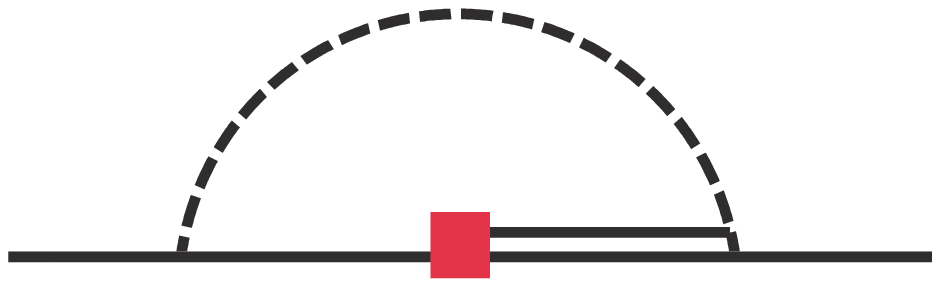,height=4cm}
\end{center}
\vspace{-.7in}
\caption{Contribution to the chiral logarithms in the 
quark distribution $\Delta u(x)-\Delta d(x)$} of 
the nucleon from the $N-\Delta$ transition distribution. 
\end{figure}

In Ref. \cite{CJ}, we have calculated the 
leading-chiral logarithms in the moments of the quark 
helicity distributions of the nucleon:  
\begin{equation}
 \langle x^{n-1}\rangle_{(\Delta u-\Delta d)N}
  = \tilde C_n\left(1-{2g_A^2+1\over (4\pi f_\pi)^2} 
  m_\pi^2\log\left({m_\pi\over\mu^2}\right) \right) \ , 
\end{equation}
where $\tilde C_n = \langle x^{n-1}\rangle_{(\Delta u-\Delta d)N}^0$. 
The delta contributions can be calculated in the same way
as shown in Ref. \cite{AS}. [An excellent introduction to
the heavy delta formalism can be found in Ref. \cite{holstein}.]
Here the contributions come in two ways:
First, there is the delta-nucleon transition 
contribution from Fig. 1: 
\begin{equation}
       - \langle x^{n-1}\rangle_{(\Delta u-\Delta d)N\Delta}^0 
      {8g_A g_{\pi N\Delta}\over 9(4\pi f_\pi)^2}
           J_2(\Delta,m_\pi) \ ,
\end{equation}
where 
\begin{eqnarray}
  J_2(\Delta, m_\pi)
  = && (3m_\pi^2-2\Delta^2)\log\left({m_\pi^2\over
    \mu^2}\right) + {2\pi m_\pi^3\over \Delta} \nonumber \\
 && + 2{(\Delta^2-m_\pi^2)^{3/2}\over \Delta}
\log\left(\Delta-\sqrt{\Delta^2-m_\pi^2+i\epsilon}
       \over \Delta + \sqrt{\Delta^2- m_\pi^2+i\epsilon}
\right) \ .
\end{eqnarray} 
Then there is the contribution from the quark-helicity 
distributions in the delta: 
\begin{equation}
        \langle x^{n-1} \rangle_{(\Delta u-\Delta d)\Delta^+} {g_{\pi N\Delta}^2\over 
              (4\pi f_\pi)^2}{100\over 9}J_1(\Delta, m_\pi)
      -  \langle x^{n-1} \rangle_{(\Delta u-\Delta d)N} 
         {4g_{\pi N\Delta}^2\over 
              (4\pi f_\pi)^2}J_1(\Delta, m_\pi) \ . 
\end{equation}
The above results are consistent with the one-loop chiral
corrections to the axial vector coupling constant $g_A$ \cite{bernard}. 
Assuming the large-$N_c$ relations,
\begin{eqnarray}
      \langle x^{n-1}\rangle_{(\Delta u-\Delta d)N\Delta}
   &=& \sqrt{2} \tilde C_n \ , \nonumber \\
 \langle x^{n-1}\rangle_{(\Delta u-\Delta d)\Delta^+}
   &=& {1\over 5} \tilde C_n \ , 
\end{eqnarray}
we find
\begin{eqnarray}
 \langle x^{n-1}\rangle_{(\Delta u-\Delta d)N}
   & = & \tilde C_n\left(1-{2g_A^2+1\over (4\pi f_\pi)^2} 
  m_\pi^2\log\left({m_\pi^2\over\mu^2}
   \right) \right. \nonumber \\
   && \left. -{g_A^2\over (4\pi f_\pi)^2}\left[-{4\over 3}
      J_2(\Delta,m_\pi)+2J_1(\Delta,m_\pi)\right]\right) \ .
\end{eqnarray}
Using the physical values of $\Delta$ and $m_\pi$, 
we find that the delta contribution cancels about 60\%
of the nucleon contribution.  
A similar result applies for the transversity moments:
\begin{eqnarray}
 \langle x^{n-1}\rangle_{(\delta u-\delta d)N}
  = && \overline{C}_n\left(1-{2g_A^2+1/2\over (4\pi f_\pi)^2} 
  m_\pi^2\log\left({m_\pi^2\over\mu^2}
   \right) \right. \nonumber \\
   && \left. -{g_A^2\over (4\pi f_\pi)^2}\left[-{4\over 3}
      J_2(\Delta,m_\pi)+2J_1(\Delta,m_\pi)\right]\right) \ .
\end{eqnarray}
where $\overline{C}_n =  \langle x^{n-1}\rangle_{(\delta u-\delta d)N}^0$
is the moment in the chiral limit.

\vspace{0.2in}
4. {\bf Vector-isoscalar channel}:

Finally, we consider the isoscalar $\Delta u(x)+ \Delta d(x)$ distribution. 
The twist-two operators ${\cal O}^{(n)i}_{\Delta u+\Delta d}$ 
are proportional to the angular momentum operator $J^i$. 
Due to the cancellation of the different spin orientations,
their matrix elements in the nucleon and delta states
are of order $N_c^{1-n}$ in $N_c$ counting. 
We can make the following expansion of the operators,
\begin{equation}
     {\cal O}_{\Delta u+\Delta d}^{(n)i}/N_c^{1-n}  
          = a_0^{(n)} 
     J^i + a_1^{(n)} {G^{ij} T^j \over N_c} + 
       a_2^{(n)} {J^2J^i\over N_c^2} + \cdots  \ . 
\end{equation}
Therefore, to order $1/N_c^2$, the distributions in the
nucleon and $\delta$ are the same,
\begin{equation}
     \Delta u_N(x) + \Delta d_N(x) 
     = \left(\Delta u_\Delta(x)+ 
      \Delta d_\Delta(x)\right)\left(1 + {\cal O}(1/N_c^2)\right) \ .
\label{vs}
\end{equation}

We have calculated the delta contribution to the chiral logarithms
in the moments of the isoscalar quark helicity distribution,
\begin{equation}
\delta\langle x^{n-1}_{(\Delta u+\Delta d)N} \rangle_{{\rm from} ~\Delta}
=
     -{4g_{\pi N\Delta}^2\over (4\pi f_\pi)^2} J_1(\Delta, m_\pi)
        \left(\langle x_{(\Delta u+\Delta d)N}^{n-1}\rangle^0 -{5\over 3}
          \langle x_{(\Delta u+\Delta d)
        \Delta}^{n-1}\rangle^0\right) \ , 
\end{equation} 
which is identical to the scalar-isovector case.
Assuming the large-$N_c$ relation, we have the sum
of the $\pi$-delta and $\pi$-nucleon loop contributions,
\begin{equation}
  \delta\langle x^{n-1}_{(\Delta u-\Delta d)N} \rangle
= \langle x_{(\Delta u+\Delta d)N}^{n-1}\rangle^0  {1\over (4\pi f_\pi)^2}
  \left[3g_A^2 J_1(\Delta,m_\pi) -
3g_A^2 m_\pi^2\log\left({m_\pi^2\over \mu^2}
   \right)\right] \ ,
\end{equation} 
which is subleading in $N_c$ counting.      

Combining the relation in Eq. (\ref{vs}) with that from 
Eq. (\ref{vv}), we find 
\begin{eqnarray}
  \Delta u_\Delta (x) &=& \left[{5+2T_z\over 10}\Delta u_N 
  + {5-2T_z\over 10}\Delta d_N \right]\left(1+ {\cal O}(1/N_c^2)\right) \ , 
  \nonumber \\
   \Delta d_\Delta (x) &=& \left[{5-2T_z\over 10}\Delta u_N 
+ {5+2T_z\over 10}\Delta d_N \right]\left(1+ {\cal O}(1/N_c^2)\right) \ . 
\end{eqnarray}
In particular, for $\Delta^+$, we have $\Delta u_\Delta
=(3\Delta u_N+ 2\Delta d_N)/5$ and $\Delta d_\Delta
=(2\Delta u_N+ 3\Delta d_N)/5$. The same relation 
exits between the transversity distributions in the
nucleon and delta.

One final comment: Since the chiral correction to the
quark distributions in the nucleon is at relative order
$1/N_c$ and since the relationships between the 
distributions in the delta and the nucleon are accurate
up to relative order $1/N_c^2$, the chiral logarithms
in the delta are connected to those in the nucleon
by the same relations above.

To summarize, we have found large-$N_c$ relations
between the quark distributions in the nucleon and 
delta, accurate to relative order $1/N_c^2$. The result
can be used to make estimates of the 
delta contributions to the chiral logarithms
in the quark distributions in the nucleon. Other uses
of the quark distributions in the delta are also
discussed briefly.

\acknowledgements
We wish to thank T. Cohen for useful discussions
about large-$N_c$ physics and M. V. Polyakov for his
constructive comments and references. We acknowledge the 
support of the U.S.~Department of Energy under 
grant no. DE-FG02-93ER-40762.

\end{document}